\documentstyle[seceq,epsf]{ptptex}

\title{
Euclidean Algorithm for a Gravitational Lens 
in a Polynomial Equation
}

\author{
Hideki {\sc Asada},$^{1,2,}$
\footnote{E-mail: asada@phys.hirosaki-u.ac.jp}
Taketoshi {\sc Kasai}$^{1}$ 
and 
Masumi {\sc Kasai}$^{1,}$\footnote{E-mail: 
kasai@phys.hirosaki-u.ac.jp}
}

\inst{
$^1$ 
Faculty of Science and Technology, Hirosaki University, 
Hirosaki 036-8561, Japan\\
$^2$ 
GReCO, Institute for Astrophysics at Paris, 
98bis boulevard Arago, 75014 Paris, France
}

\recdate{
April 7, 2004
}

\abst{
The Euclidean algorithm in algebra is applied to a class of 
gravitational lenses for which the lens equation consists of 
any set of coupled polynomial equations in the image position.  
In general, this algorithm allows us to reduce an apparently 
coupled system to a single polynomial in one variable 
(say $x$ in Cartesian coordinates) 
without the other component (say $y$), which is expressed as 
a function of the first component. 
This reduction enables us to investigate the lensing properties 
in an algebraic manner: For instance, we can obtain 
an analytic expression of the caustics by computing 
the discriminant of the polynomial equation. 
To illustrate this Euclidean algorithm, we re-examine a binary 
gravitational lens and show that the lens equation is reduced to 
a single real fifth-order equation, in agreement with previous works. 
We apply this algorithm also to the linearlized Kerr lens 
and find that the lens equation is reduced to 
a single real fifth-order one. 
}

\begin{document}

\maketitle

\section{Introduction}
Gravitational lensing plays an important role 
in modern astronomy, \cite{SEF} in particular in the problems of 
determining the cosmological parameters, probing the mass profile 
of galaxies, and detecting a dark object such as MACHO. 
When the lens mass distribution is axially symmetric 
along the line of sight, the basic equation is reduced to 
one dimension. 
Except in this case, the system is coupled nonlinearly in general. 
It is thus believed that we need numerical treatments. 
Recently, however, the lens equation, which is apparently coupled, 
has been shown by chance to be reduced to a single real algebraic 
equation even for asymmetric cases, such as those of a two-point 
mass \cite{Asada02} and an isothermal ellipsoid. \cite{AHK} 
It should be noted that the lens equation can be expressed as 
a single {\it complex} algebraic equation in the complex formalism. 
\cite{Witt90,Witt93}
However, this equation is nonlinearly coupled in the real and 
imaginary parts, and therefore we must solve it numerically. 

The reduction to a single real algebraic equation enables us 
in principle to study the lensing properties analytically or 
numerically in an efficient and accurate manner. 
For instance, we can obtain an analytic expression for the 
caustics, \cite{AKK,AHK} which is a curve in the source plane, 
and a criteria for a change in the number of images. 
The reason that this can be done is that this change 
corresponds to that in the number of real roots for the single 
real algebraic equation, which is determined by a discriminant 
in algebra. \cite{Waerden} 

It is worthwhile mentioning the resultant method, a part of 
the elimination theory in algebra. \cite{Waerden,WM,Dalal} 
This method gives us a condition that two polynomial 
equations have a common root. 
For instance, let us take the Cartesian coordinates $(x, y)$ 
to denote these equations by $P(x, y)=0$ and $Q(x, y)=0$. 
We can apply the resultant method to our problem: 
First, these equations are considered only for $y$. 
A common root $y$ for them exists if the resultant vanishes, 
where the resultant is a function of $x$, denoted by $X(x)$.
Similarly, a common root $x$ exists if for the resultant $Y(y)$, 
we have $Y(y)=0$, if we pay attention to only $x$. 
However, all of the roots for these equations, $X(x)=0$ and 
$Y(y)=0$, do not necessarily satisfy the original equations 
$P(x, y)=0$ and $Q(x, y)=0$: 
An appropriate combination of a root $x$ for $X(x)=0$ 
and $y$ for $Y(y)=0$ is true, but the resultant method 
itself never tells about a way of choosing it. 

In the examples mentioned above, the binary \cite{Asada02} and 
the isothermal ellipsoid, \cite{AHK} 
we have a one-dimensional equation for one component of the coordinates  
and a linear equation for the other component of the coordinates, 
which is solved trivially and thus regarded as a function. 
In this sense, such a formalism, in which we have a 
one-dimensional equation and the associated function, 
is not known for a more general lens system. 
The main purpose of this paper is to demonstrate that 
the Euclidean algorithm is widely applicable to the lens equation 
for a pair of real polynomials.
In section 2, we study a binary gravitational lens as an example. 
In the Appendix, a linearized Kerr lens is investigated 
as another example.

\section{Euclidean algorithm}
\subsection{Binary Gravitational Lens}
We consider a binary system consisting of two point masses, 
$M_1$ and $M_2$, and a separation vector $\mbox{\boldmath $L$}$ 
from object 1 to 2, which is located at a distance 
$D_{\mbox{L}}$ from the observer.
In units of the Einstein ring radius angle 
$\theta_{\mbox{E}}$, the lens equation reads \cite{Asada02,AKK} 
\begin{equation}
\mbox{\boldmath $\beta$}=\mbox{\boldmath $\theta$}
-\Bigl( 
\nu_1 \frac{\mbox{\boldmath $\theta$}}{|\mbox{\boldmath $\theta$}|^2} 
+\nu_2 \frac{\mbox{\boldmath $\theta$}
-\mbox{\boldmath $\ell$}}{|\mbox{\boldmath $\theta$}
-\mbox{\boldmath $\ell$}|^2} 
\Bigr) , 
\label{lenseq}
\end{equation}
where $\mbox{\boldmath $\beta$}=(a, b)$ and 
$\mbox{\boldmath $\theta$}=(x, y)$ denote 
the vectors representing the positions of the source and image, 
respectively, and we defined the mass ratios $\nu_1$ and $\nu_2$ and 
the angular separation vector $\mbox{\boldmath $\ell$}$ as 
\begin{eqnarray}
\nu_1&=&\frac{M_1}{M_1+M_2} , \\
\nu_2&=&\frac{M_2}{M_1+M_2} , \\
\mbox{\boldmath $\ell$}&=&
\frac{\mbox{\boldmath $L$}}{D_{\mbox{L}}\theta_{\mbox{E}}} . 
\end{eqnarray}
We have the identity $\nu_1+\nu_2=1$. 
For brevity, $\nu_2$ is denoted by $\nu$. 
Equation (\ref{lenseq}) consists of a set of two coupled real quintic
equations for $(x, y)$ 
\begin{equation}
|\mbox{\boldmath $\theta$}|^2 
|\mbox{\boldmath $\theta$}-\mbox{\boldmath $\ell$}|^2 
(\mbox{\boldmath $\theta$}-\mbox{\boldmath $\beta$})
-(1-\nu) |\mbox{\boldmath $\theta$}-\mbox{\boldmath $\ell$}|^2 
\mbox{\boldmath $\theta$}
-\nu |\mbox{\boldmath $\theta$}|^2 
(\mbox{\boldmath $\theta$}-\mbox{\boldmath $\ell$})=0 . 
\label{lenseq2}
\end{equation}

For two point masses at different distances, the lens equation 
becomes more complicated. This has been previously investigated 
(e.g., Erdl and Schneider (1993) \cite{ES}). 

\subsection{Euclidean algorithm}
First, let us briefly review the Euclidean algorithm 
for integers. \cite{Waerden} 
For example, we search for a common factor of $20$ and $12$. 
The larger integer, $20$, is divided by the smaller one, $12$, 
with the remainder $8$, and $12$ is divided by $8$ 
with the remainder $4$. Such a procedure is repeated until 
the remainder vanishes. 
Thus, we obtain 
\begin{eqnarray}
20&=&1\times 12+8 , \nonumber\\
12&=&1\times 8+4 , \nonumber\\
8&=&2\times 4+0 , 
\label{integers}
\end{eqnarray}
which shows that $4$ is the greatest common divisor (G.C.D.)  
of $20$ and $12$. 
Actually, Eq. ($\ref{integers}$) can be rearranged as 
\begin{eqnarray}
20&=&5\times 4+0 ,\nonumber\\
12&=&3\times 4+0 . 
\label{integers2}
\end{eqnarray}
As another example, consider $9$ and $4$. 
In this case, we obtain 
\begin{eqnarray}
9&=&2\times 4+1 , 
\nonumber\\
4&=&4\times 1+0 . 
\end{eqnarray}
Thus unity is the G.C.D. in this case. 
This procedure is called the Euclidean algorithm, which is 
applicable widely in algebra. 
Let us apply it to a pair of polynomials, given in 
Eq. ($\ref{lenseq2}$), in which a common factor gives us a common root, 
namely an image position. 

Equation ($\ref{lenseq2}$) is coupled with respect to $x$ and $y$. 
For simplicity, the separation of the binary is assumed to be 
on the $x$-axis $\mbox{\boldmath $\ell$}=(\ell, 0)$. 
Concerning $x$, Eq. ($\ref{lenseq2}$) consists of 
a fifth-order equation and a fourth-order equation, respectively:  
\begin{eqnarray}
f_5&\equiv&\sum_{i=0}^5 c_i x^i=0 , 
\label{f5}
\\
f_4&\equiv&\sum_{i=0}^4 d_i x^i=0 . 
\label{f4}
\end{eqnarray}
Here, all of the coefficients are polynomials 
in $y$, $a$, $b$, $\ell$ and $\nu$:  
\begin{eqnarray}
c_5&=&1 , \\
c_4&=&-(a+2\ell) , \\
c_3&=&2y^2+\ell^2+2a\ell-1 , \\
c_2&=&-2(a+\ell)y^2+\ell(2-a\ell-\nu) , \\
c_1&=&y^4+(\ell^2+2a\ell-1)y^2-\ell^2(1-\nu) , \\
c_0&=&-ay^4+\ell y^2(\nu-a\ell) , \\
d_4&=&y-b , \\
d_3&=&-2\ell y+2b\ell , \\
d_2&=&2y^3-2by^2+(\ell^2-1)y-b\ell^2 , \\
d_1&=&-2\ell y^3+2b\ell y^2+2\ell(1-\nu)y , \\
d_0&=&y^5-by^4+(\ell^2-1)y^3-b\ell^2y^2-\ell^2(1-\nu)y . 
\end{eqnarray} 
We divide $f_5$ by $f_4$ with the quotient $q_3$ and 
the remainder $f_3$ as 
\begin{equation}
f_5=q_3 f_4+f_3 , 
\end{equation}
where $f_3$ is cubic in $x$. 
This procedure can be repeated as 
\begin{eqnarray}
f_4&=&q_2 f_3+f_2 , 
\nonumber\\
f_3&=&q_1 f_2+f_1 , 
\nonumber\\
f_2&=&q_0 f_1+f_0 , 
\end{eqnarray}
where $f_i$ is $i$-th order in $x$ and $q_i$ is linear in $x$. 
We can eliminate $f_3$ and $f_2$ from these equations 
in order to show that both $f_5$ and $f_4$ have 
a common factor $f_1$ if $f_0$ vanishes.  
In short, Eqs. ($\ref{f5}$) and ($\ref{f4}$) are equivalent 
to the pair of equations $f_0=0$ and $f_1=0$. 
In the cases of integers considered above, 
Eqs. ($\ref{integers}$) and ($\ref{integers2}$), 
$f_1$ and $f_0$ correspond to $4$ and $0$, respectively. 

Because $f_1$ is linear in $x$, it takes the form 
\begin{equation}
f_1=\Bigl(D(y) x+E(y)\Bigr) G(y) , 
\end{equation}
where $D(y)$ and $E(y)$ are polynomials in $y$, $a$, $b$, 
$\ell$ and $\nu$, and $G(y)$ is an algebraic function. 
Because $G(y)$ plays no role in the following discussion, 
we write $D(y)$ and $E(y)$ as 
\begin{eqnarray}
D(y)&=&b^3 \ell^2 - 2 b^3 \ell^2 \nu + 2 b^3 \ell^2 \nu^2 
\nonumber\\
&&+ (-2 b^2 - 4 a b^2 \ell + 4 b^2 \ell^2 
+ a^2 b^2 \ell^2 + b^4 \ell^2 + 8 a b^2 \ell \nu 
\nonumber\\
&&- 10 b^2 \ell^2 \nu - 2 a b^2 \ell^3 \nu + b^2 \ell^4 \nu 
+ 6 b^2 \ell^2 \nu^2)y 
\nonumber\\
&&+ (-2 a^2 b - 6 b^3 - 4 a^3 b \ell - 4 a b^3 \ell 
+ 3 a^2 b \ell^2 + 3 b^3 \ell^2 + 4 a b \ell \nu 
\nonumber\\
&&+ 8 a^3 b \ell \nu + 8 a b^3 \ell \nu - 8 b \ell^2 \nu 
- 12 a^2 b \ell^2 \nu - 4 b^3 \ell^2 \nu + 6 a b \ell^3 \nu 
\nonumber\\
&&- b \ell^4 \nu + 6 b \ell^2 \nu^2)y^2 
\nonumber\\
&&+ (2 a^2 + 2 b^2 - 4 a^2 b^2 - 4 b^4 + 4 a^3 \ell 
+ 4 a b^2 \ell - 4 a^2 \ell^2 - 4 b^2 \ell^2 
\nonumber\\
&&- 4 a \ell \nu - 8 a^3 \ell \nu + 12 a^2 \ell^2 \nu 
- 4 a \ell^3 \nu + 2 \ell^2 \nu^2)y^3 
\nonumber\\
&&+ (4 a^2 b + 4 b^3 - 8 a b \ell \nu + 4 b \ell^2 \nu)y^4 , 
\label{D}
\end{eqnarray}
\begin{eqnarray}
E(y)&=&-b^3 \ell^3 + 2 b^3 \ell^3 \nu - b^3 \ell^3 \nu^2 
\nonumber\\
&&+ (b^2 \ell + 3 a b^2 \ell^2 - 3 b^2 \ell^3 
- a^2 b^2 \ell^3 - b^4 \ell^3 - 4 a b^2 \ell^2 \nu 
\nonumber\\
&&+ 6 b^2 \ell^3 \nu + a^2 b^2 \ell^3 \nu + b^4 \ell^3 \nu 
- 3 b^2 \ell^3 \nu^2)y 
\nonumber\\
&&+ (a^2 b \ell + b^3 \ell + 3 a^3 b \ell^2 
+ 3 a b^3 \ell^2 - 2 a^2 b \ell^3 
\nonumber\\
&&- 2 b^3 \ell^3 + 4 b^3 \ell \nu - 2 a b \ell^2 \nu 
- 4 a^3 b \ell^2 \nu - 4 a b^3 \ell^2 \nu + 4 b \ell^3\ \nu 
\nonumber\\
&&+ 4 a^2 b \ell^3 \nu + 2 b^3 \ell^3 \nu - a b \ell^4 \nu 
- 3 b \ell^3 \nu^2)y^2 
\nonumber\\
&&+ (4 a b^2 - a^2 \ell - 5 b^2 \ell 
- 3 a^3 \ell^2 - 3 a b^2 \ell^2 + 3 a^2 \ell^3 + 3 b^2 \ell^3 
\nonumber\\
&&+ 4 b^2 \ell \nu + 4 a^2 b^2 \ell \nu + 4 b^4 \ell \nu 
+ 2 a \ell^2 \nu + 4 a^3 \ell^2 \nu - 4 a b^2 \ell^2 \nu 
\nonumber\\
&&- 5 a^2 \ell^3 \nu + b^2 \ell^3 \nu + a \ell^4 \nu 
- \ell^3 \nu^2)y^3 
\nonumber\\
&&+ (4 a^3 b + 4 a b^3 - 4 a^2 b \ell 
- 4 b^3 \ell - 8 a^2 b \ell \nu + 12 a b \ell^2 \nu 
- 4 b \ell^3 \nu)y^4 
\nonumber\\
&&+ (-4 a^3 - 4 a b^2 
+ 4 a^2 \ell + 4 b^2 \ell + 4 a^2 \ell \nu - 4 b^2 \ell \nu 
- 4 a \ell^2 \nu)y^5 . 
\end{eqnarray} 

The $x$ component of an image position is determined by 
\begin{equation}
x=-\frac{E(y)}{D(y)} , 
\label{x}
\end{equation} 
if the component $y$ is given. 
The $y$ component is obtained by solving $f_0=0$, 
which does not depend on $x$. 
After factorization, this is reduced to a fifth-order 
equation in $y$,  
\begin{equation}
\sum_{i=0}^5 e_iy^i=0 , 
\label{fifth}
\end{equation}
where
\begin{eqnarray}
e_5&=&-4(a^2 + b^2)(a^2 + b^2 - 2 a \ell + \ell^2) , 
\\
e_4&=&4 b\Bigl(-a^2 + a^4 - b^2 + 2 a^2 b^2 + b^4 - 2 a^3 \ell 
- 2 a b^2 \ell + a^2 \ell^2 + b^2 \ell^2 
\nonumber\\
&&+ (2 a \ell - \ell^2) \nu \Bigr) , 
\\
e_3&=&-a^2 - b^2 + 8 a^2 b^2 + 8 b^4 - 2 a^3 \ell 
- 10 a b^2 \ell + 2 a^2 \ell^2 - a^4 \ell^2 + 6 b^2 \ell^2 
\nonumber\\
&&- 2 a^2 b^2 \ell^2 - b^4 \ell^2 + 2 a^3 \ell^3 + 2 a b^2 \ell^3 
- a^2 \ell^4 - b^2 \ell^4 
\nonumber\\
&&+ (2 a \ell + 4 a^3 \ell + 4 a b^2 \ell - 6 a^2 \ell^2 
- 2 b^2 \ell^2 + 2 a \ell^3) \nu - \ell^2 \nu^2 , 
\\
e_2&=&\Bigl(a^2 + 5 b^2 + 2 a^3 \ell + 2 a b^2 \ell - 2 a^2 \ell^2 
+ a^4 \ell^2 - 2 b^2 \ell^2 + 2 a^2 b^2 \ell^2 + b^4 \ell^2 
\nonumber\\
&&- 2 a^3 \ell^3 - 2 a b^2 \ell^3 + a^2 \ell^4 + b^2 \ell^4 
\nonumber\\
&&+ (-2 a \ell - 4 a^3 \ell - 4 a b^2 \ell + 4 \ell^2 
+ 6 a^2 \ell^2 + 2 b^2 \ell^2 - 2 a \ell^3)\nu 
\nonumber\\
&&- 3 \ell^2 \nu^2\Bigr) , \\
e_1&=&1 + 2 a \ell - 2 \ell^2 + a^2 \ell^2 + b^2 \ell^2 
- 2 a \ell^3 + \ell^4 
\nonumber\\
&&+ (-4 a \ell + 5 \ell^2 + 2 a \ell^3 - \ell^4)\nu - 3 \ell^2 \nu^2 , 
\\
e_0&=&-b^3 \ell^2 (-1 + \nu)\nu . 
\end{eqnarray}

The common root of Eqs. ($\ref{f5}$) and ($\ref{f4}$) is 
thus obtained by (1) solving Eq. ($\ref{fifth}$) for $y$ and 
(2) substituting a root $y$ into Eq. ($\ref{x}$) to determine $x$.

\subsection{Particular cases}
There can exist multiple roots in algebraic equations. 
A condition for a true multiple root is a discriminant in algebra, 
which corresponds to the caustics in the theory of 
gravitational lenses. 
An apparent multiple root is due to the adopted coordinate system: 
For instance, let us imagine two roots with the same value 
of $y$ but different $x$. 
These two images appear on a line parallel to the $x$-axis. 
A rotation of the coordinate system breaks such an apparent 
multiplicity. 
Here, we discuss this apparent multiplicity in our algorithm. 

In the preceding subsection, we have assumed $f_i\neq 0$. 
However, $f_1=0$ may be satisfied automatically 
for a particular set of the lens and source parameters.  
Then, a common factor for Eqs. ($\ref{f5}$) and ($\ref{f4}$) 
becomes $f_2$, which is second order in $x$. 
Hence, apparent multiple roots are located at the same $y$ but 
different $x$. 
In this particular case, a practical strategy for solving 
the lens equation in our algorithm is 
(1) solving Eq. ($\ref{fifth}$) for $y$ to find single roots 
and multiple roots, 
(2a) substituting a single root $y$ into Eq. ($\ref{x}$) 
to determine $x$, and 
(2b) substituting a multiple root $y$ into $f_2=0$ 
and solving it as a square equation for $x$. 
If the multiplicity is three, the multiple root $y$ must 
be substituted into $f_3=0$, which becomes a cubic equation for $x$. 

Let us consider the example $(a, b)=(1/2, 1/4)$ 
for $\ell=1$ and $\nu=1/2$. 
Then, the fifth-order equation is factorized as  
\begin{equation}
(2 + 5 y)^2 (-1 - 12 y - 4 y^2 + 16 y^3) = 0 . 
\end{equation}
For its multiple root $y=-2/5$, $f_2$ becomes 
\begin{equation}
f_2=-\frac{123}{50000}(25 x^2-25 x-16) . 
\end{equation}
Hence $f_2=0$ is solved as $x=(5\pm\sqrt{89})/10$.

\section{Conclusion}
We demonstrated that the Euclidean algorithm works well 
in the case of a binary gravitational lens: 
The image position for Eqs. ($\ref{f5}$) and ($\ref{f4}$), 
which are a set of coupled polynomial equations,  is 
obtained by (1) solving Eq. ($\ref{fifth}$) for $y$ and 
(2) substituting a root $y$ into Eq. ($\ref{x}$) to determine $x$. 
In the case of multiple roots, we must substitute 
the multiple root $y$ into $f_2=0$ 
and solve it as a square equation for $x$. 
If the multiplicity is three, the multiple root $y$ must 
be substituted into $f_3=0$, which becomes a cubic equation for $x$. 

The Euclidean algorithm is applicable to any pair of polynomials. 
The linearized Kerr lens is investigated in the Appendix. 
It would be interesting to investigate $N$ point masses or some 
analytic models for a galaxy \cite{SEF} 
if the lens equation is a polynomial. 
These interesting applications will be considered in the future.

\section*{Acknowledgements}
We would like to thank H. Nakazato and D. Tambara 
for fruitful conversations. 
One of the authors (H. A.) would like to thank L. Blanchet 
for his hospitality at the institute for Astrophysics at Paris, 
where this work was done. 
This work was supported in part (H. A.) by a fellowship 
for visiting scholars from the Ministry of Education of Japan.

\appendix
\section{A Linearized Kerr Lens} 
We consider the linearized Kerr spacetime with mass $m$ and 
specific angular momentum vector $\mbox{\boldmath $q$}$. 
In units of the Einstein ring angular radius $\theta_{\mbox{E}}$, 
the lens equation for this spacetime is expressed as \cite{AK,AKY}
\begin{equation}
\mbox{\boldmath $\beta$}=\mbox{\boldmath $\theta$}
-\frac{\mbox{\boldmath $\theta$}}{|\mbox{\boldmath $\theta$}|^2} 
+\left(
\frac{\mbox{\boldmath $s$} \times \mbox{\boldmath $\bar e$}}
{|\mbox{\boldmath $\theta$}|^2}
-\frac{2 \{(\mbox{\boldmath $s$} \times \mbox{\boldmath $\bar e$})
\cdot \mbox{\boldmath $\theta$} \} \mbox{\boldmath $\theta$} }
{|\mbox{\boldmath $\theta$}|^4}
\right) , 
\end{equation}
where $\mbox{\boldmath $\bar e$}$ denotes the unit vector 
along the line of sight, and we have defined 
\begin{equation}
\mbox{\boldmath $s$}
=
\frac{D_{\mbox{S}} \theta_{\mbox{E}}}{4 D_{\mbox{LS}}}
\frac{\mbox{\boldmath $q$}-(\mbox{\boldmath $q$} 
\cdot \mbox{\boldmath $\bar e$})\mbox{\boldmath $\bar e$}}{m} . 
\end{equation}
Here, $D_{\mbox{S}}$ and $D_{\mbox{LS}}$ denote the distance 
between the observer and the source and the distance 
between the lens and the source, respectively. 
Although the orientation of the spin vector $\mbox{\boldmath $q$}$ is 
arbitrary, only the projection of $\mbox{\boldmath $q$}$ onto the lens
plane makes contribution. 
On the lens plane, we adopt the Cartesian coordinates 
such that $\mbox{\boldmath $s$}=(0, s)$, 
$\mbox{\boldmath $\beta$}=(a, b)$ and 
$\mbox{\boldmath $\theta$}=(x, y)$. 
The lens equation becomes fifth-order and fourth-order in $x$, 
respectively as 
\begin{eqnarray}
g_5 &\equiv& (x-a)(x^2+y^2)^2+(s-x)(x^2+y^2)-2sx^2 = 0 , 
\label{g5}\\
g_4 &\equiv& (y-b)(x^2+y^2)^2-y(x^2+y^2)-2sxy = 0 , 
\label{g4}
\end{eqnarray}
which are coupled nonlinearly. 

By applying the Euclidean algorithm, the pair of polynomial 
lens equations are reduced to 
\begin{equation}
x=-\frac{H(y)}{I(y)} , 
\end{equation}
\begin{equation}
\sum^{5}_{i=0}k_i y^i=0 , 
\end{equation}
where we have defined 
\begin{eqnarray}
H(y)&=&2 b^3 s y^2 + 2 b^2 y^3 (a + s + a^2 s + b^2 s)
\nonumber\\
&&+ 2 a b y^4 (a^2 + b^2 - 2 a s) 
+ 2 y^5 (-a^3 - a b^2 + a^2 s - b^2 s) , \\
\label{H}
I(y)&=&b^3 s^2 + b^2 y (-1 + 4 a s + 3 s^2) 
+ b y^2 (-a^2 - 3 b^2 + 2 a s + 4 a^3 s + 4 a b^2 s + 3 s^2) 
\nonumber\\
&&+ y^3 (a^2 + b^2 - 2 a^2 b^2 - 2 b^4 - 2 a s - 4 a^3 s + s^2) 
\nonumber\\
&&+ 2 b y^4 (a^2 + b^2 - 2 a s) , 
\label{I}
\end{eqnarray}
and 
\begin{eqnarray}
k_5&=&4(a^2 + b^2)^2 , 
\label{k5}\\
k_4&=&-4 b (-a^2 + a^4 - b^2 + 2 a^2 b^2 + b^4 + 2 a s) , 
\label{k4}\\
k_3&=&a^2 + b^2 - 8 a^2 b^2 - 8 b^4 - 2 a s (1 + 2 a^2 + 2 b^2) 
+ s^2 , 
\label{k3}\\
k_2&=&b [-a^2 - 5 b^2 + 2 a s (1+ 2 a^2 + 2 b^2) + 3 s^2] , 
\label{k2}\\
k_1&=&b^2 (-1 + 4 a s + 3 s^2) , 
\label{k1}\\
k_0&=&b^3 s^2 . 
\label{k0}
\end{eqnarray}
Thus the lens equation is reduced to a single, real fifth-order one.

\end{document}